# Nonequilibrium dynamics and thermodynamics provide the underlying physical mechanism of the perceptual rivalry

Yuxuan Wu[1], Liufang Xu[1*], Jin Wang[2**]

[1]Biophysics & Complex System Center, Center of Theoretical Physics, College of Physics, Jilin University, Changchun 130012, People's Republic of China
[2]Department of Chemistry and of Physics and Astronomy, Stony Brook University, Stony Brook, New York 11794, United States
*Correspondence: lfxuphy@jlu.edu.cn
**Correspondence: jin.wang.1@stonybrook.edu

Perceptual rivalry, where conflicting sensory information leads to alternating perceptions crucial for associated cognitive function, has attracted researcher's attention for long. Despite progresses being made, recent studies have revealed limitations and inconsistencies in our understanding across various rivalry contexts. We develop a unified physical framework, where perception undergoes a consecutive phase transition process encompassing different multi-state competitions. We reveal the underlying mechanisms of perceptual rivalry by identifying dominant switching paths among perceptual states and quantifying mean perceptual durations, switching frequencies, and proportions of different perceptions. We uncover the underlying nonequilibrium dynamics and thermodynamics by analyzing average nonequilibrium flux and entropy production rate, while associated time series irreversibility reflects the underlying nonequilibrium mechanism of perceptual rivalry and link thermodynamical results with neuro-electrophysiological experiments. Our framework provides a global and physical understanding of brain perception, which may go beyond cognitive science or psychology but embodies the connection with wider fields as decision-making.

## INTRODUCTION

Perceptual rivalry, the phenomenon where conflicting sensory information leads to alternating perceptions, has captivated researchers for nearly two centuries [1,2]. Initially confined to visual perception, it now encompasses various sensory modalities, including tactile [3,4], olfactory [5], and auditory domains [6,7], , and even interactions between different senses [8,9]. Although interpreted as a form of bistable switching traditionally, perceptual rivalry was enriched by the discovery of "interocular grouping" in 1996 revealing the coexistence of multiple perceptions within the rivalry process [10]. This finding, supported by numerous subsequent studies [11-17], underscores the non-unique nature of perceptual dynamical phases in rivalry. Meanwhile, since their introduction in 1965, Levelt's four propositions have served as foundational principles in binocular rivalry research [18]. These propositions describe the influence of stimulus strength on perceptual dominance, dominance duration, and alternation rate. However, recent studies have revealed limitations and inconsistencies in their applicability across various rivalry contexts [19-22].

Underlying common mechanisms of varying perceptions emerges from all these findings. Different sensory stimuli as neuroelectrical signals are finely processed and transmitted to different senior cortices based on their specific neural pathways [23,24]. The interaction dynamics at the macroscopic level between their corresponding neural populations can share striking similarities. This resonates with the holistic concept of brain function [25,26]. Following the appropriate 'order parameter' changes (e.g., brightness in binocular rivalry or frequency difference in auditory stream segregation), varying perceptions can undergo similar phase transition process. This highlights the need for a unified framework for understanding. Given the established connection between perceptual switching and brain decision-making [27,28], such a framework would be of significant value.

Several theoretical frameworks have been proposed to understand rivalry dynamics [16,29-32]. Early models, emphasizing adaptation as the switching mechanism, depicted perceptual states alternating driven by noise [31,33-35]. While adaptation, representing neural fatigue, undoubtedly plays a role, such models often predict regular switching patterns inconsistent with experimental observations, necessitating the inclusion of noise for rectification. However, recent findings suggest a more prominent and dominant role for noise in shaping rivalry dynamics [16,36-39]. Building upon these insights, we propose a neurodynamical explanation of perceptual rivalry in which noise drives perceptual switching, with adaptation serving as an auxiliary mechanism.

This fluctuating nature of perceptual rivalry requires a framework capable of capturing its dynamic energy exchanges with the environment: nonequilibrium physics emerges as a powerful tool for this purpose [40-49]. We employ this framework to develop a unified model encompassing five distinct phases: (1) no obvious perception, (2) weakly excited perceptions, (3) stable and dominant perceptual bistability, (4) perceptual tristability induced by strong stimuli, and (5) fully mixed perception under intense stimuli. Phase transitions connect these phases, suggesting that seemingly incompatible phenomena like bistable and multistable rivalry (recently observed in visual competition [50]) can coexist within the same system under varying conditions.

We first quantify the global potential landscape and the probability in the steady state of the system [51]. Combining them with stochastic dynamical simulations, we calculate the barrier height and mean duration time of each perceptual state, offering insights into the dynamics of perceptual switching and corresponding to the duration of perceptual excitation observed in experiments. We also identify dominant paths for state transitions, quantifying the underlying process and aligning with experimental observations. Furthermore, average nonequilibrium flux and entropy production rate (*EPR*) reveal the thermodynamical underpinnings of the phase transitions. We demonstrate their close connection to time irreversibility, reflecting the degree of detailed balance breaking. These global properties shed light on the physical mechanisms governing the progression through different perceptual phases and transitions, particularly explaining the prevalence of binary rivalry in many perceptual rivalry scenarios. Experimentally, these findings can be readily investigated by analyzing time sequences of perceptual signals using techniques like EEG, ECoG, and fMRI [52-57], offering insights into the global cost. Finally, we extend our model to encompass situations where the stimuli or inhibitions on both sides are not equal and analyze the binocular imbalance [58,59]. Further, in this work, we analyze the predictions from our model based on Levelt's propositions, offering a comprehensive discussion. Through this comparison, we aim to provide novel insights into the generalizability and potential limitations of these well-established principles.

## RESULTS
### The dynamical model and global landscapes under varying external stimulation

We start with a model shown in Fig.1 [36], which generalizes the biological mechanism in vision or hearing. The model comprises four key components: excitatory populations "*X*" and "*Y*" for the rivaling perceptions, inhibitory populations X_inh and Y_inh, a collective excitatory population "Col_pop", and external stimuli "$g_x$" and "$g_y$". Notably, rivalry emerges indirectly through Col_pop, rather than direct mutual inhibition.

Focusing on the mean firing rates of the excitatory population *X* denoted by $x$ and *Y* denoted by $y$, the dynamical function is determined as:

$$\frac{dx}{dt} = F_x = -\gamma x + \sigma + \Lambda(x) * f[\alpha x + g_x - \beta_x (\phi_{col\_inh} + \phi^x_{self\_inh})^2]$$
$$\frac{dy}{dt} = F_y = -\gamma y + \sigma + \Lambda(y) * f[\alpha y + g_y - \beta_y (\phi_{col\_inh} + \phi^y_{self\_inh})^2]$$
(1)

in which $f(x) = 1/\{1 + \exp[(x-\theta)/k]\}$ is a sigmoid function representing neuronal input-output. The three terms within the sigmoidal function represents: Self-activation $\alpha x$ or $\alpha y$ with a strength coefficient $\alpha$, external stimuli $g_x$ or $g_y$, and inhibition with coefficients $\beta_x$ or $\beta_y$. The inhibition term includes $\phi_{col\_inh} = \{1 + \tanh(10[q(x+y) + g_x + g_y])\}$, the mean firing rate of X_inh or Y_inh with a weight coefficient $q$, and self-inhibition, $\phi^x_{self\_inh} = \eta x$, $\phi^y_{self\_inh} = \eta y$ with a weight coefficient $\eta$. $\Lambda(x) = \{1 - a_1(1 + \tanh[(x-c)/a_2])\}$ represents adaptation by an inverted hyperbolic tangent function, capturing the decreased neuronal reactivity with sustained stimulation [60]. Additionally, $\sigma$ incorporates external factors like deformation [61], electromagnetic stimulation [62] and bioluminescent photon emission [63] that can influence neuronal activation beyond direct signal transduction. See Appendix for details of all parameters. Eq.(1) describes the interactions between different neural populations in the neural pathway by the driving force ***F***, which lead to the formation of complex perceptual neural functions.

We focus on the external stimuli $g_x$ and $g_y$, which can represent various factors in different rivalry modalities, such as color [64], brightness [65] or contrast ratio [66] in binocular rivalry, auditory stream modes in auditory rivalry [67], or even ambiguous vibration patterns in tactile rivalry [4]. Notably, $g_x$ and $g_y$ can go beyond physical intensity, encompassing feature values like streaky pattern angles in binocular rivalry [50]. Numerical solutions of Eq.(1) reveal five deterministic dynamical phases (Fig.2(**a1-h1**)), representing different perceptual rival states. These phases are separated by four phase transitions or bifurcations of $g_x$ and $g_y$: K1 ($g_x = g_y = 0.009$), K2 ($g_x = g_y = 0.011$), K3 ($g_x = g_y = 0.0177$), and K4 ($g_x = g_y = 0.021$). Fig.2(**a1-h1**) also demonstrates the continuity of these phases, where any rivalry state corresponds to a specific point in the phase plane. Employing Eq.(2) in the section of methods, the one-dimensional stochastic trajectories of states switching are visualized (Fig.2(**a3-h3**)).

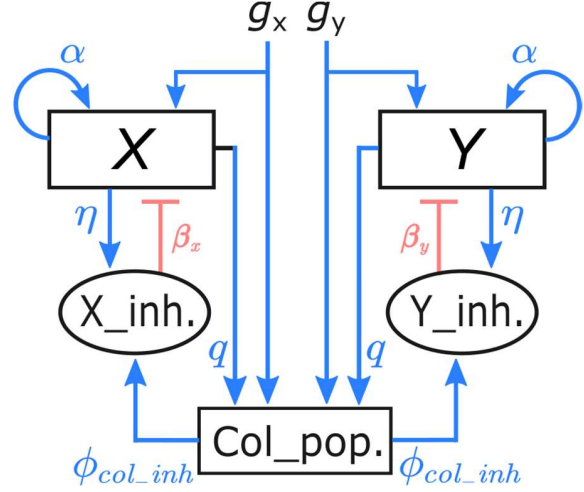

**Figure 1. Circuit structure of bilateral perception**
"*X*" and "*Y*" represents two excitatory neural populations respectively. "X_inh." is the inhibitory neural population of the *X* perception and "Y_inh." is the inhibitory neural population of *Y*. "Col_pop." is a collective inhibitory neural population, which is excited by external stimuli "$g_x$" and "$g_y$" and two excitatory populations. The excitatory projections are colored blue, and the inhibitory projections are colored red. The coefficients in Eq.(1) are tagged beside the corresponding pathways.

Landscape and flux, introduced in the section of methods, can depict the global nonequilibrium dynamics and thermodynamics of the system arising from neural circuit interactions. Without these interactions, all possible states would have equal probability. However, neural interactions generate a probability landscape where certain states hold higher probabilities (basins of attractors) representing biologically significant functional states. These basins (dark blue areas in Fig.2(**a2-h2**, **a4-h4**)) correspond to exciting perceptions *X*, *Y*, or a mixed state *M*. Barriers separate these basins, and noise drives the system across these barriers, leading to state transitions. Noise amplification can flatten the landscape, reducing the effective barrier heights, while weakening noise has the opposite effect. Consequently, landscape barriers naturally measure the global stability of states, reflecting the difficulty of switching between them. Moreover, barrier height strongly correlates with state stability, which translates to the duration of a particular perception (Fig.2(**b4-g4**)).

On the other hand, $\bm{J}_{ss}$ as the fluxes represent the flow of the probability between states in the nonequilibrium steady state. These flows, often diverging from the landscape gradient (Fig.2), act as additional driving forces influencing state transitions. Unlike the gradient $-\nabla U$ which is along with $\bm{F}$ largely, $\bm{J}_{ss}$ can dynamically destabilize the current state and lead to new ones. Thus, this can lead to dynamical instability of the current state and the emergence of the possible new states. This "tearing apart" effect contributes to the dynamical origins of instability, bifurcations, and nonequilibrium phase transitions. Based on these insights, we explore the specific characteristics of each dynamical phase.

**Binary rivalry**. This typical phase, marked as "*Bi-Riv*", occurring within the range $g_x = g_y \in (0.011, 0.0177)$, exhibits classic bistable behavior. It possesses two stable states (*X* and *Y*) representing dominant perceptions (Fig.2(**d1**, **e1**)), where a dominant side suppresses a weaker one. In binocular or olfactory rivalry, *X* could signify the dominant left-eye or left-nostril perception, while *Y* signifies the right (Fig. 3(**a1**, **a3**)). Similarly, for ambiguous figures (Fig.3(**a2**)), *X* could represent the dominant vase perception, while *Y* represents the dominant faces. This analogy extends naturally to other bistable perceptual competitions.

Stochastic trajectories depict the rivalry and transitions between the two dominant states (Fig.2(**d3**, **e3**)). Fig.2(**d2**, **e2**) shows two landscape basins separated by a barrier, corresponding to *X* and *Y*. Vortices of the rotational $\bm{J}_{ss}$ flux from two sides connect on the barrier, manifesting the states switching.

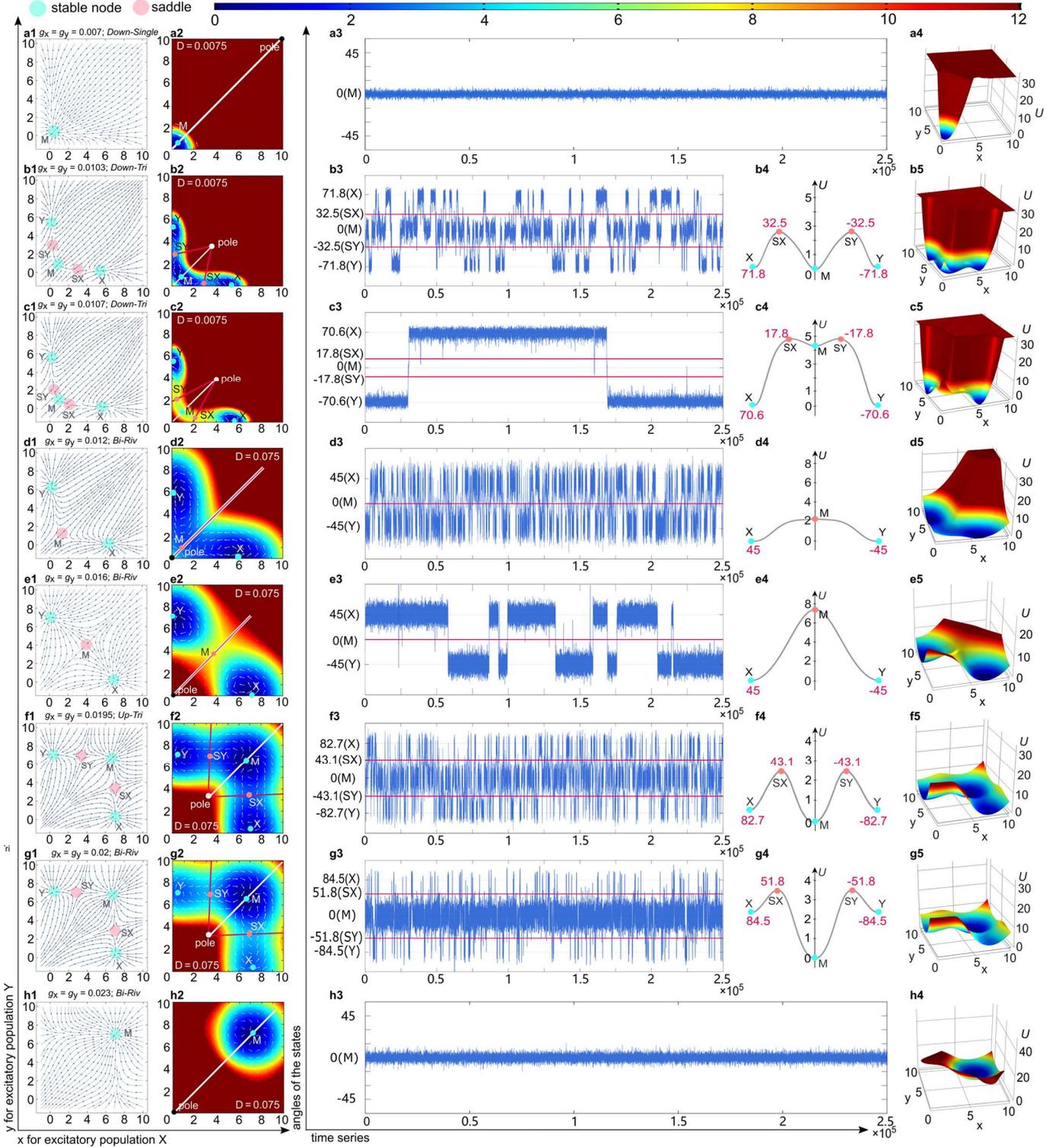

**Figure 2. Deterministic dynamical structure, landscapes, stochastic one-dimensional trajectories and sections of barriers on the landscapes**

Subfigures in each row possesses a same set of coefficients. (**a1-h1**): Deterministic dynamical structures of every dynamic phase, displayed as deterministic stream fields. (**a2-h2**): Vertical views of landscapes. All of these subgraphs (**a2-h2** and the 3D views **a5, h5, b5-g5**) have an identical color range. White arrows represent fluxes $J_{ss}$, sizes of which depict where there are significant probability flows. For the convenience of portraying, we take the logarithm of the real $J_{ss}$ values. (**a3-h3**): One-dimensional stochastic trajectories. Labels of the ordinate are angles of the state points at a certain position on the landscape during a simulation, and the label of abscissa are the time of the stochastic simulation. Note that in (**b4-g4**), connecting lines of the fixed points are selected as the sections of the landscape barriers. (**a2-h2**) also show the method to measure the angle. For the monostable phases, the poles locate at (0, 0) or (10, 10). For the tristable phases, the poles locate at the centers of the circles determined by *X*, *Y*, *M*. For the bistable phase, the abscissa of *X* and the ordinate of *Y* determine the position of the pole. The polar axes (gray lines) are from the poles to the point *M*. Thus, the angle between a state and the polar axis can be obtained. The magenta lines from the poles to the saddles separate different areas of the perceptual states.

**Ternary rivalry**. This phase, named as "*Up-Tri*", occurs within $g_x = g_y \in (0.0177, 0.021)$. It exhibits that not only can one side dominate, but both sides can combine into a new mixed perception in varying degrees (Fig.3(**b1-b3**)). The deterministic dynamics reveals that the fixed point *M*, previously representing a saddle, migrates diagonally and becomes a stable state, while two new saddle points *SX* and *SY* emerge (Fig.2(**f, g**)). This stable *M* state embodies the mixed perception arising from simultaneous excitation of *X* and *Y* by external signals [9,12,50]. Notably, the "interocular grouping" phenomenon, demonstrating coherence between both eyes during rivalry, aligns with this case [10]. Fig.2(**f3, g3**) confirms the alternating dominance among three perceptual states.

The landscape now features three basins of attraction corresponding to *X*, *Y* and *M*, representing the unilateral and mixed perceptions, respectively. Interestingly, the $J_{ss}$ flux suggests that transitions between unilateral states *X* and *Y* are most likely mediated by the mixed state M (Fig.2(**f2, g2**)). We will explore the underlying reason in the following section.

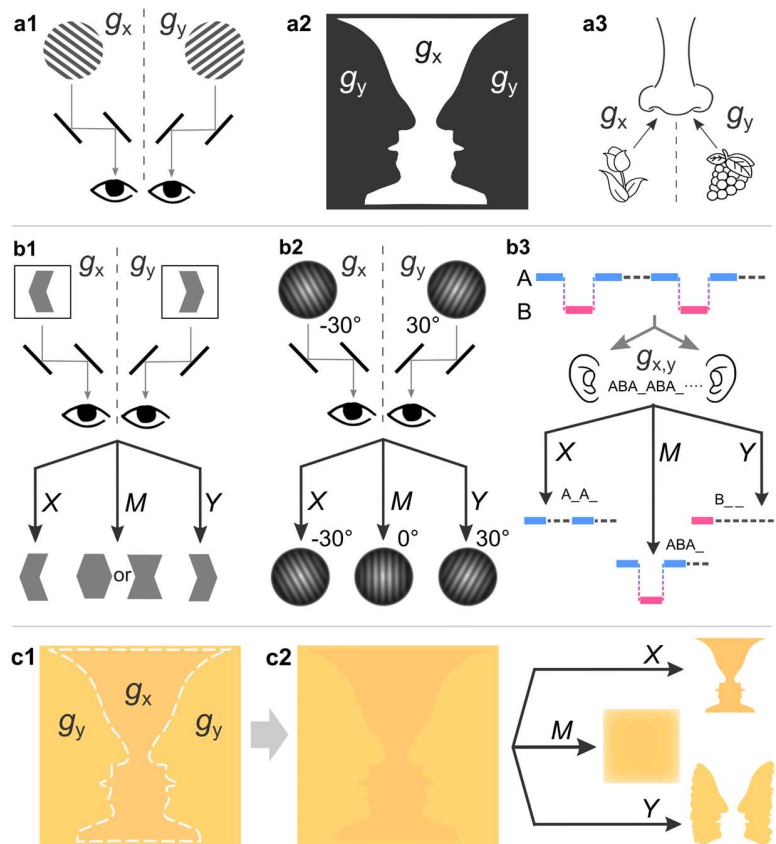

**Figure 3. Illustrations of varying perception rivalries**
(**a1-a3**): Typical cases of the bistable perceptions. Perceptions *X*, *Y* accord with stimulations $g_x$, $g_y$. (**a1**): Classic binocular rivalry of streaky patterns. Using the stereoscope, two different patterns are individually input to the left eye or the right eye as stimulations. (**a2**): Famous example of the visual rivalry between ambiguous figures: Rubin's vase, first proposed by Edgar Rubin in 1912. (**a3**): Olfactory rivalry between bistable states. Generally, this kind of rivalry are generated by using two different smells to stimulate the left or right nostrils respectively. (**b1-b3**): Typical multistable perceptions. (**b1**) shows that when showing herringbone patterns respectively to a single eye, tester will possibly see one of those two mixed patterns besides herringbone patterns of a single dominant eye [12]. (**b2**) shows a tristable visual perception formed by two stripe patterns with opposite angles to the vertical direction [50]. (**b3**) shows the classic mode of auditory stream segregation [68]. (**c1, c2**): A simple experiment done by the authors. Note that in (**c1**) the vase and faces are painted in different but similar colors. In practical operation, the two colors are much closer than they are currently. Subgraph (**c2**) is identical with (**c1**) except the dash line is eliminated. Other colors and chroma contrasts are also tested by authors. Subjects will perceive the vase, faces, or only a square with unified color without intentionally distinguishing the patterns.

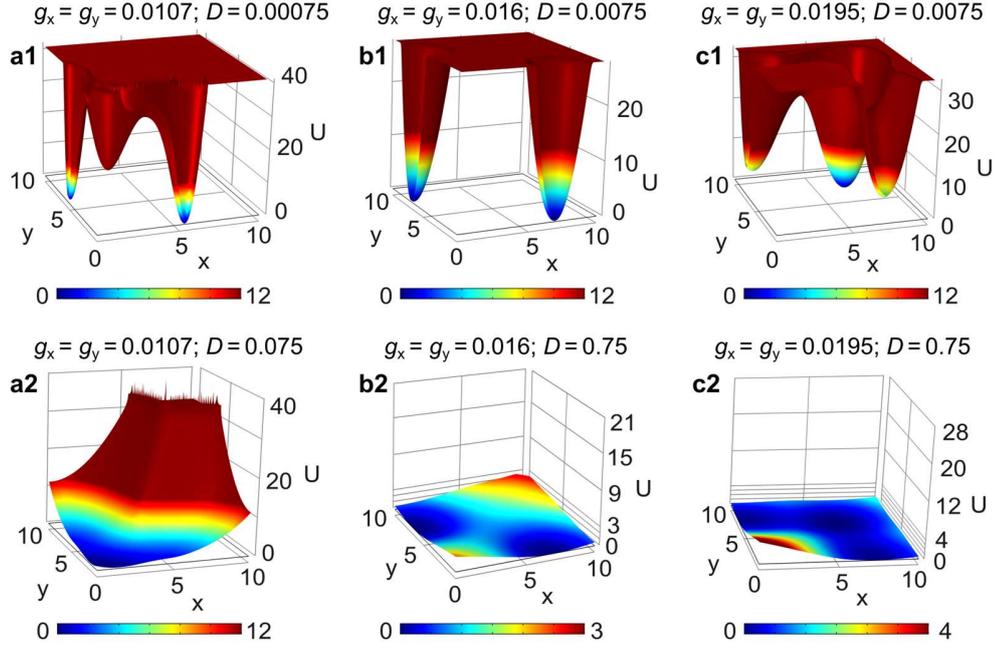

**Figure 4. Impact of inappropriate noise intensity**
Subgraphs (**a1**, **a2**), (**b1**, **b2**) and (**c1**, **c2**) respectively display the landscapes of Down-Tri, Bi-Riv and Up-Tri phases. Comparing with Fig.2(**c, e, f**), (**a1, b1, c1**) are based on the tenfold reduced $D$ values, while the $D$ values of (**a2, b2, c2**) are magnified tenfold.

**Quiescent state**. This monostable phase, denoted as "*Down-Single*," occurs within the parameter range $g_x = g_y \in (0.005, 0.009)$ (Fig.2(**a1**)). It is characterized by a single attractor $M$ at the bottom left of the phase plane. This state signifies the quiescence of both $X$ and $Y$ populations due to weak external stimulation. Notably, this phase often serves as the initial stage when external stimulation increases in various perceptual rivalries.

The stochastic trajectories corroborate this quiescent behavior (Fig.2(**a3**)). The landscape also reflects this, with a single basin located at bottom left on the phase plane (Fig.2(**a2**)). Interestingly, the rotational $J_{ss}$ flux forms two symmetric vortices within this basin (Fig. 2a2), aligning with the dynamics evolving (Fig.2(**a2**)).

**Over exciting state**. This phase, denoted as "*Up-Single*", occurs within the range $g_x = g_y \in (0.021, 0.025)$. It represents a situation where both $X$ and $Y$ populations are persistently excited and do not return to unilateral states. For instance, consider the case of Fig.3(**b2**): When the angles between two presented stripe patterns are very close to 0°, the subjects can almost observe only the pattern $M$, as shown in reference [50]. However, prolonged maintenance of such exclusive mixed perceptions is rarely reported [69,70], due to its low importance in the experiments.

Stochastic trajectories (Fig.2(**h3**)), landscape, and $J_{ss}$ (Fig.2(**h2**)) exhibit the corresponding behaviors of this phase. Interestingly, the $J_{ss}$ resembles the inverted version of the *Down-Single* phase. As $g_x$ and $g_y$ decrease, the $J_{ss}$ vortices elongate towards the unilateral states, eventually transitioning to the *Up-Tri* phase.

**Faint perceptions**. This phase, denoted as "*Down-Tri*", resides within the parameter range $g_x = g_y \in (0.009, 0.011)$. Building upon the binary rivalry structure, an additional attractor $M$ emerges as a quiescent state at the bottom left (Fig.2(**b1, c1**)). Three stable states, separated by saddles $SX$ and $SY$, lead to the possibility of transitioning from quiescence to either unilateral excited state, essentially forming a potential tristability. One-dimensional trajectories depict the switching between these states (Fig.2(**b3, c3**)). We design a simple test to verify the existence of this faint perceptions (Fig.3(**c1, c2**)).

Notably, $X$ and $Y$ positions on the axes are lower than those in the *Bi-Riv* phase. As external stimulation increases, they gradually approach *Bi-Riv* levels. This implies that *Down-Tri* represents faintly perceptible unilateral states $X$ and $Y$ that can be stochastically excited under weak stimulation. Essentially, it serves as a transitional phase between *Down-Single* and *Bi-Riv*. Although rarely reported in experiments, we delve deeper into this phenomenon from the perspectives of dissipation cost and detailed balance breaking.

Compared to other phases, the landscape and trajectory reveal distinct features. The basins of $X$, $Y$, and $M$ are noticeably smaller ((Fig.2(**b2, c2**)), indicating stronger dominance and fewer fluctuations in unilateral states. This suggests that the weak stimuli make it difficult for another state to challenge the dominant one. The $J_{ss}$ flux exhibits two vortexes condensing and expanding towards the unilateral states $X$ and $Y$, mirroring the landscape and trajectory (Fig.2(**b2, c2**)).

The appropriate noise strength is significant for the stability of perceptual switching process [71]. Note that the *Down-Single* and *Down-Tri* phases in Fig.2 uses a smaller value of $D=0.0075$ compared to other phases. For these two phases, the practical $D$ range for biological function is $[0.005, 0.0125]$, while that of other phases is $[0.05, 0.125]$. While other phases remain mostly unaffected, *Down-Tri* loses its biological function when $D$ approaches or exceeds 0.05. The interval of the practical $D$ range of *Down-Tri* is compact, reflecting its sensitivity to noise strength. Fig.4(**a-c**) exhibit the impact of inappropriate noise intensity. Apparently, insufficient $D$ leads to excessively high barriers that hinders switching within a reasonable timeframe, while excessive $D$ flattens the landscape, resulting in the unconstrained motion of the states that hinders the biophysical function. Furthermore, the biological function of a perceptual phase might be linked to temporal synchronization or phase locking among neuronal populations, potentially serving as the primary source of noise effect [72,73].

## Dominant path of perception switching

Understanding the most possible paths between perceptual states is crucial for characterizing the underlying switching process. To this end, we employ the dominant path that introduced in the Methods section. Fig.5 presents the results of this analysis.

Firstly, the dominant paths for forward and backward switching do not coincide (Fig.5). Secondly, these paths notably bypass saddles, even the mixed attractor $M$ in Fig.5(**a, c**). Thirdly, in Fig.5(**a, c**), the path from $X$ to $M$ and $M$ to $Y$ differs from the path from $X$ directly to $Y$. These observations highlight the inherent irreversibility of the global dynamics, driven by the combined interplay of the landscape gradient $-\nabla U$ and $J_{ss}$ flux. Essentially, the dominant path condenses the complex kinetic process of states switching in stochastic dynamics to a single, most possible route. This route represents the physically achievable and practical pathway for perceptual switching in the system.

Dominant path analysis determines that the switching paths between $X$ and $Y$ in the *Up-Tri* phase are intermediated by $M$ state, although shorter distance on the landscape seems to favor direct paths (Fig.5(**a, c**)). In Ref.[12], besides the switching intermediated by a mixed perception, observers also have a probability to capture the direct switching between left and right herringbone (Fig.3(**b**)). It demonstrates the fast transition between $X$ and $Y$ intermediated by $M$ basin (Fig.2(**b3, c3, f3, g3**)).

Furthermore, for *Up-Tri* phase near the bifurcation K3 ($g_x=g_y=0.0177$), $M$ basin appears shallow and metastable, which blurs the distinction between binary and ternary rivalry. For instance, when observers see the Rubin's vase (Fig.3(**a2**)), a jumbled mixed pattern serves as an intermediation during the switching between the vase and faces, although it may be hardly recognized.

In addition, some reports discuss the effect of memory in multiple perceptual competitions [50,74,75]. Whereas, our model reveals equal probabilities for transitions from $M$ to either $X$ or $Y$. This aligns with the situation of "zero coherence" in brain decision-making, where states have equal chances of switching to the correct or incorrect option [76]. Dynamics under equal external stimulation dictate this balanced decision-making within the mixed state, offering no apparent evidence of inherent memory bias.

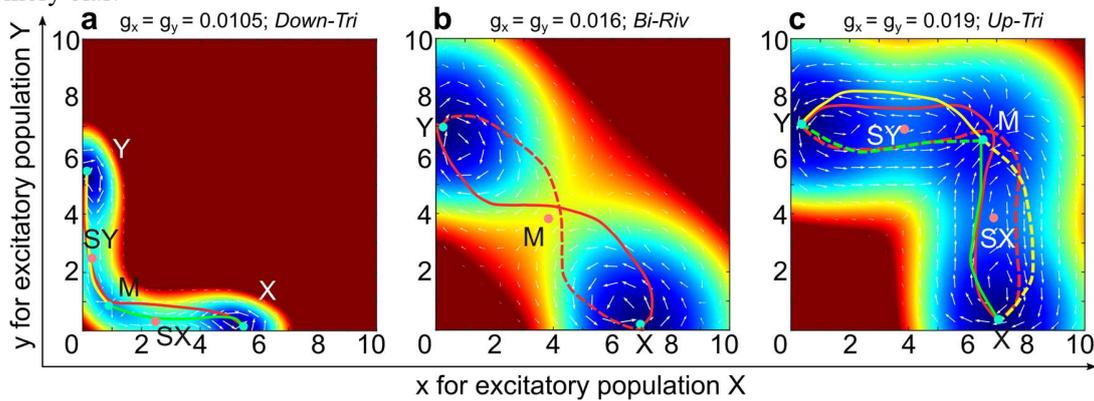

**Figure 5. Dominant paths in different dynamical phases**
In these graphs, following Fig.2, spring green points represent stable nodes, and pink points for saddles. Color scales are according to the Fig.2. In (**a-c**), red solid lines are dominant paths from the stable node of $X$ to $Y$. In (**a, c**), yellow solid lines are paths from stable nodes of the mixed state $M$ to $Y$, and green lines are paths from $X$ to $M$. For bistable system, the backward paths from $Y$ to $X$ are shown as the red dash line in (**b**). For tristable system, in (**c**), the backward path from $Y$ to $M$ and from $M$ to $X$ are respectively shown by the green and yellow dash lines, while the backward path from $Y$ directly to $X$ are shown by the red dash line. The backward paths are not displayed in (**a**) for a clear observation.

## Dynamics of perception switching: barrier height and duration time

Experimental studies often rely on readily measurable quantities like perceptual duration (or switching frequency) [66,77] and average proportion [50,78] to capture the essence of perceptual rivalry. These measures form the basis of Levelt's propositions [79]. In this section, we introduce a theoretical framework to calculate mean duration times (MDT) and barrier heights (BH) directly from the model's landscapes. Further, we ascertain that durations and proportions of different perceptions exhibit a positive correlation. These results can be validated in psychophysical experiments.

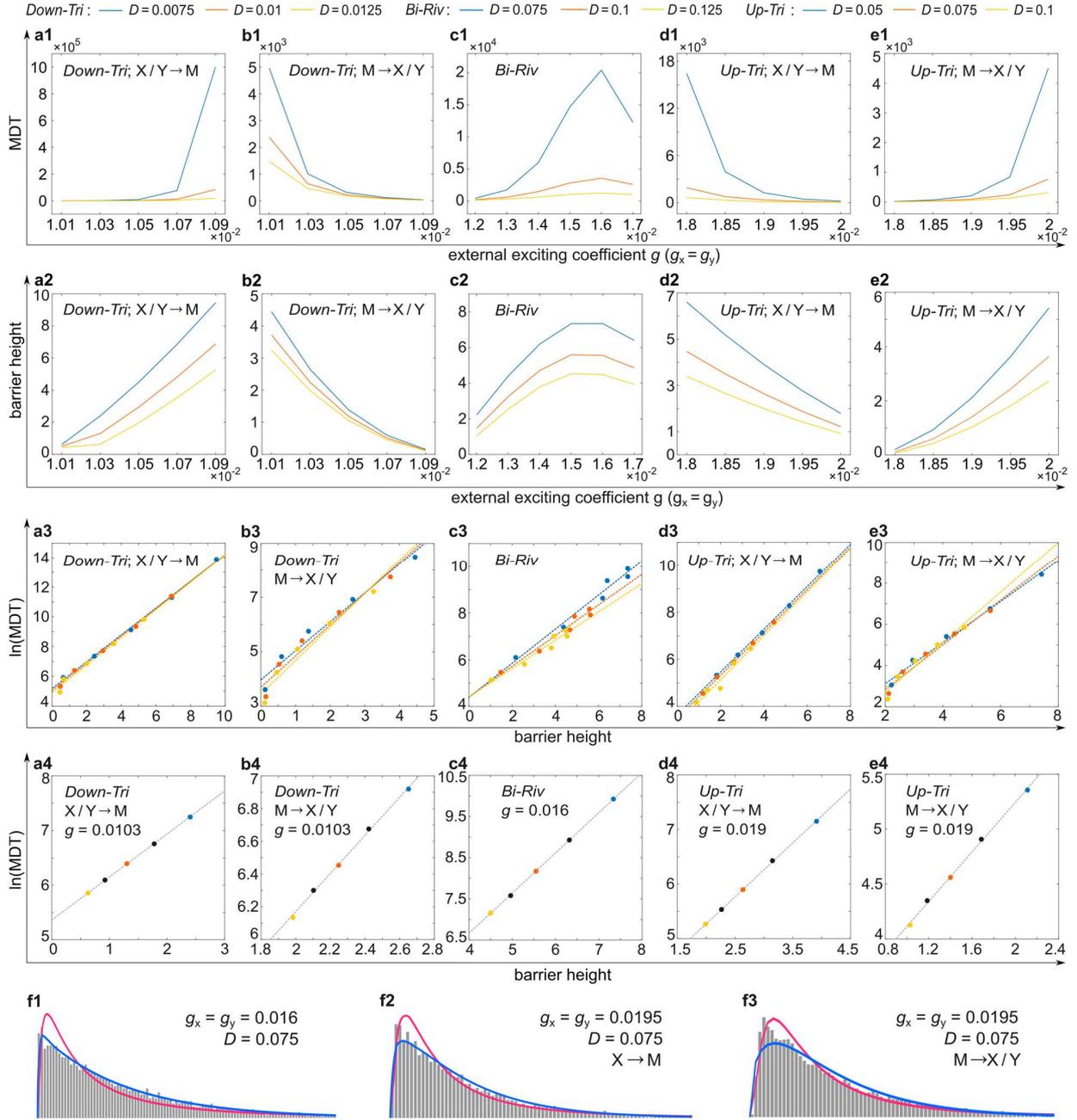

**Figure 6. MDT, BH, linear fitting curves of natural logarithm MDT values and BH, and MDT distributions**

In (**a1-e1**) and (**a2-e2**), $g$ values of the sampling points constituting polylines in subgraphs are abscissa values. (**a3-e3**): The linear fitting between the ln(MDT) and BH, in which each fitting line uses the sample points of one $D$ value but different $g$ values. (**a4-e4**): The case of one $g$ value with different values of $D$ is selected from each perceptual phase as an example, in which dash lines are fitting curves of MDT and BH. (**f1-f3**): The distribution of the single duration time, in which the pink line is the lognormal distribution fitting, and the blue line is the Gamma distribution fitting.

The single duration time (SDT) for the *Bi-Riv* phase is defined as the time a state takes for originating from the *X* or *Y* attractors to reach the connecting line between the pole and the saddle (polar axis in Fig.2(**d2**, **e2**)). This definition extends to the *Up-Tri* and *Down-Tri* phases for the mixed state *M*. Here, the SDT is the escape time from *M* to either connecting line between the pole and saddles *SX* or *SY* (red polar axes in Fig.2(**b2**, **c2**, **f2**, **g2**)). Additionally, the SDT for *X* or *Y* is defined as the time to reach the polar axes through *SX* or *SY*, respectively. In our work, the mean duration time (MDT) for a given parameter set ($D$, $g_x$, $g_y$) is calculated by averaging 10000 SDT realizations. Notably, our MDT definitions largely align with the well-established concept of mean first passage time (MFPT) (See Ref.[80] for more details.)

BH plays a crucial role in governing state switching durations. It is defined as the depth difference between attractor and saddle, largely reflects the integral of deterministic force **F** that a state needs to overcome to escape a basin (Fig.2(**c3**, **e3**)). Thus, higher noise strength $D$ facilitates overcoming **F** that represents neural circuit interactions that push states towards attractors, flattening the global landscape with the barriers, while less noise translates to fewer fluctuations, making states more resilient and leading to longer durations (more stable perceptions). For instance, if an observer sees Rubin's vase (Fig.3(**a2**)), the perceptions of vase or faces will endure for a longer duration when the barrier in binary rivalry landscape is higher. The landscape barriers directly reflect the stability of perceptions, as observed in the consistency between BH and MDT trends (Fig.6(**a1-e1**, **a2-e2**)).

In fact, our definition of BH and MDT originates from the mean first passage time as the first-order approximation, exhibiting a positive linear relationship between $D$ and the $e$ exponential of the potential difference [81]. Fig.6(**a4-e4**) demonstrate the excellent fit between MDTs and BHs for different $D$ at fixed $g$. On the other hand, changing $g$ results in the alteration of the dynamical structure. However, although the fluctuation of sampling points appears, the exponential relationship remains robust (Fig.6(**a3-c3**)). Evidently, the exponential relationship between MDT and BH is widely effective.

Although earlier studies described the distributions of dominant state duration using gamma distribution [79,82], recent works suggest lognormal distribution is also suitable [83]. As shown in Fig.6(**f1-f3**), both types of distributions fit our simulated data well, supporting the validity of our BH and MDT definition. Furthermore, this characteristic distribution reflects the nonequilibrium nature of the system [84].

Interestingly, Fig.6(**c1**) reveals a distinct peak in BH at $g_x = g_y = 0.016$. This suggests the existence of an optimal external stimulation that simultaneously reinforces both *X* and *Y* perceptions, leading to their increased stability. This finding differs from Levelt's original proposition IV for binocular rivalry, which suggested a positive correlation between stimulus strength and alternation rate [18]. However, it aligns more closely with modern interpretations that propose a negative correlation between alternation rate and stimulus strength at low stimulation levels [78]. This supports the growing recognition that rivalry dynamics are spatiotemporal, influenced by the specific range of external stimulation. Our work provides further evidence for this notion, suggesting that an "optimum" stimulation regime might exist for maximizing the persistence of both perceptions.

Fig.7(**a1**, **b1**) displays the state dominance proportions, another frequently employed experimental measure. Intuitively, in tristable phases, the proportion of *M* appears linearly related to the proportion of MDT of *M* (*M* transitioning to *X* or *Y*, divided by the sum of MDT and *X* or *Y* transitioning to *M*) [80]. This is partially supported by Fig.7(**a2**, **b2**). However, experimental studies often report weak correlations between durations and proportions [85]. This discrepancy arises from two factors: (1) Rapid alternations that states may rapidly switch back and forth within a single perception (bistability or tristability) or switch between perceptions (tristability), being difficult for observers to perceive [86]. (2) Barrier relaxation that relaxation processes occur at the barriers, which might not be readily discernible in experiments due to the coarse-grained nature of observed time series. Consequently, experiments may miss these details, leading to a blurred relationship between durations and proportions. However, simulations capture these features, as reflected in Fig.7(**a2**, **b2**), potentially offering a more accurate representation of the underlying dynamics.

While Fig.7(**a2**, **b2**) suggests a strong linear trend, statistical fluctuations are expected in finite sequences. Notably, the most significant deviations occur for *Down-Tri* ($g_x = g_y = 0.0103$) and *Up-Tri* ($g_x = g_y = 0.019$ or $0.0195$) conditions (Fig.7(**a3**, **b3**)). In these cases, the barriers between states (*X* or *Y* to *M* and vice versa) are relatively low, and the basins of *X*, *Y*, and *M* have similar weights (Fig.2(**b4**, **f4**)). These factors lead to more frequent relaxations between states, making the MDT values more susceptible to small perturbations, particularly when they are close. Conversely, in other conditions with significant disparity in MDT values, deviations are less pronounced. This significant deviation serves as a feature of rivalry between equally weighted perceptions. Experimentally, we recommend combining psychophysical tests with neuroelectrophysiological methods for conditions where different perceptions exhibit similar durations. This integrated approach could prove valuable in elucidating the underlying principles of perceptual alternation.

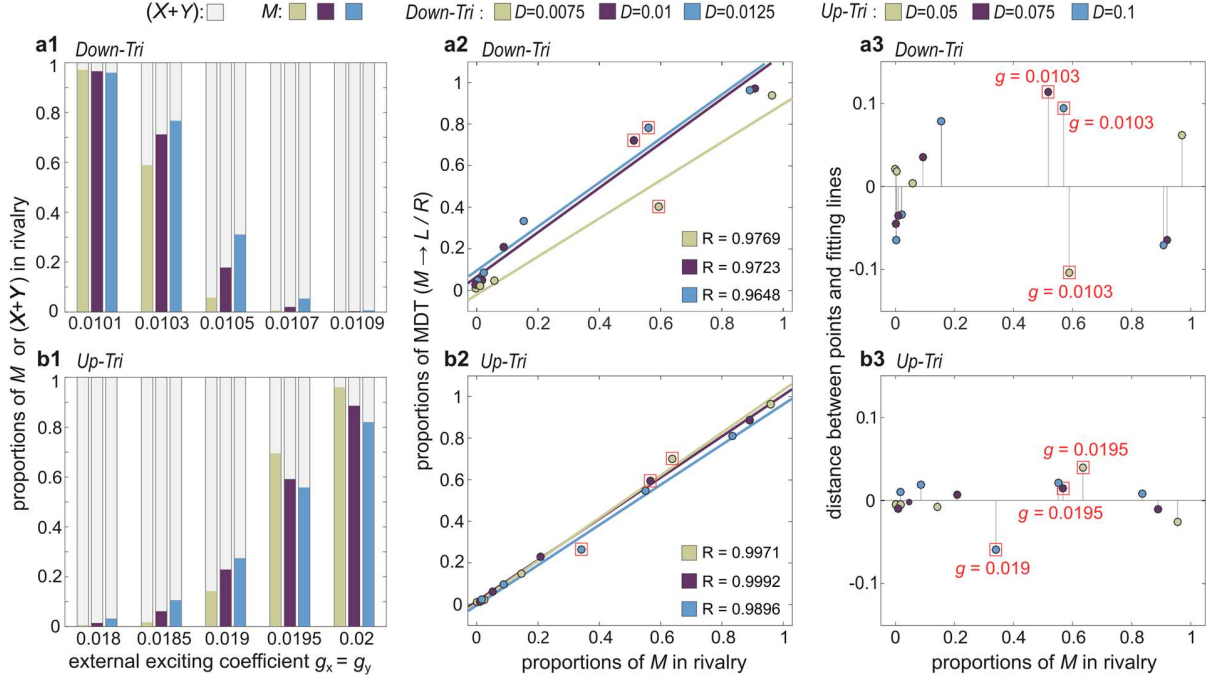

**Figure 7. Correlation between MDTs and proportions of perceptions**
(**a1, b1**): Proportions of the mixed state *M* and the unilateral states (*X+Y*) in the time series of the perceptual rival processes. For each set of $g_x$, $g_y$ and $D$, a stochastic trajectory of $2\times 10^7$ entire time steps is calculated and then the amount of steps is sum of which the state is in the domain of *M*. The definition of MDT and BH is continuously used to determine the domain of *M*. A proportion of *M* are gain as the ratio of the amount of steps that the state is in *M* domain to the total time steps. Naturally, a proportion of (*X+Y*) equals to (1-*M* proportion). (**a2, b2**): Results of the linear fitting between the MDT proportions $(M \to X \text{ or } Y)/((M \to X \text{ or } Y)+(X \text{ or } Y \to M))$ and the proportions of *M* in the time series. The linear correlation coefficients R are also presented. (**a3, b3**): Distances from the sampling points to the fitting lines. Note that the points encircled by red squares are ones whose distances are largest among all for the same $D$ values.

## Global nonequilibrium dynamics and thermodynamics: flux, entropy production rate, and time irreversibility

**Average $J_{ss}$.** Moving beyond individual states, we delve into the system's global nonequilibrium thermodynamics. As shown in Eq.(4), $J_{ss}/P_{ss}$ represents the nonequilibrium rotational force within the overall driving force. To quantify the system's global flux strength and its relation to detailed balance breaking, we employed the average $J_{ss}$ across different conditions (Eq. 5).

Fig.8(**a1**, **a2**) presents the average $J_{ss}$ results. Notably, a subtle bulge emerges in the range $g_x=g_y\in(0.0177,0.021)$ with a peak at $g_x=g_y\simeq 0.019$, which remains along with $D$ increasing. This indicates that tristable phases (*Down-Tri* and *Up-Tri*) require more dynamical support compared to monostable or bistable phases. Interestingly, the *Down-Tri* phase exhibits a remarkably high average $J_{ss}$ peak at low $D$ values, rapidly declining to near-baseline levels when $D$ exceeds 0.005. This suggests a nonequilibrium thermodynamical reason for the rarity of *Down-Tri* phase in perceptual rivalries: Maintaining the *Down-Tri* phase requires significantly more $J_{ss}$, and its practical $D$ range for biological function does not overlap with other phases.

The average flux results also offer insights into nonequilibrium phase transitions and bifurcations. In Fig.8(**a1**, **a2**), for $D=0.0075$, the *Down-Tri* phase exhibits an ascent starting around $g=0.009$, followed by a sharp rise at $g=0.0103$ and a subsequent decline around $g=0.011$. For the *Up-Tri* phase, the slight bulge starts at about $g=0.0177$ and ends at about $g=0.021$. These starting and ending points coincide with the deterministic dynamical bifurcations. Remarkably, the *Down-Tri* and *Up-Tri* peaks ($g=0.0103$ and $g=0.0195$) correspond to states where the *X*, *Y*, and *M* basins in the landscapes have nearly equal weights. In contrast, the *Bi-Riv* phase exhibits its lowest average $J_{ss}$ at $g\simeq 0.016$, where the barrier is highest. As $g$ increases towards the tristable phase range, the rotational flux strengthens. This increased flux is prone to rotate around the system rather than localize at a specific state, effectively tearing apart the current attractor and facilitating the formation of new ones. With three states, a stable flow cycle requires more $J_{ss}$ compared to one or two states. This rotational flux serves as the thermodynamic driving force for both state instability and the emergence of new states, ultimately explaining the thermodynamic basis of bifurcations and nonequilibrium phase transitions.

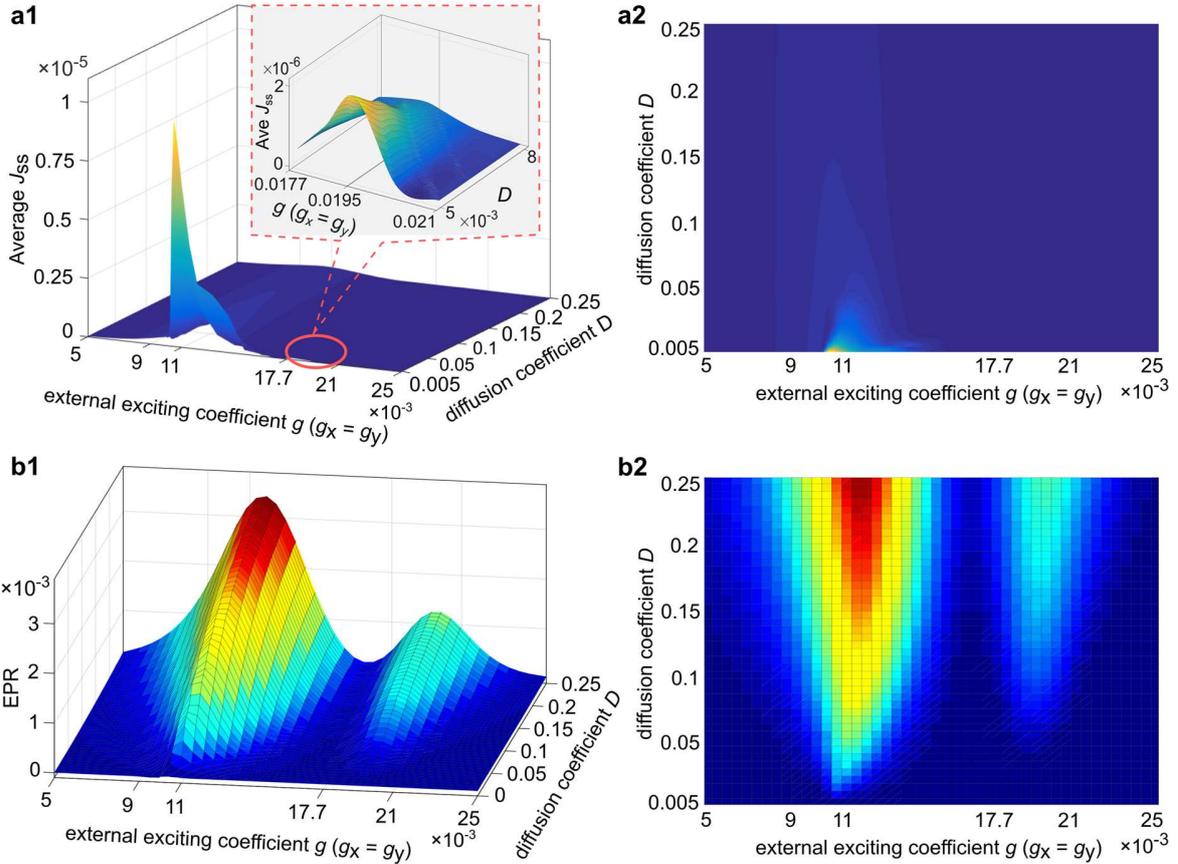

**Figure 8. Results of average flux $J_{ss}$ and entropy production rate**
Subfigure (**a1**, **b1**) is the vertical view of (**a2**, **b2**). The color ranges of (**a1**) with (**a2**) and (**b1**) with (**b2**) are identical. The trend of average flux along with the increase of $D$ monotonically decreases for the same $g$ value, while the opposite is true for the case of the *EPR*. See [87] for more interpretation.

**Entropy production rate and heat dissipation rate.** As we interpret in the Methods section, the cost of the system can be measured by the entropy production rate (*EPR*) and heat dissipation rate (*HDR*), and they are also related to the degree of the irreversibility of the system.

Fig.8(**b1**, **b2**) reveal the *EPR* results. Undulating trends with two ups and downs emerge as external stimulation $g$ increases, suggesting four significant dynamical phase transitions. Around $D=0.1$, upward or downward trends align closely with deterministic dynamical bifurcations K3 and K4 ($g=0.0177, 0.021$). Similarly, the bifurcation K1, K2 ($g=0.009, 0.011$) manifests around $D=0.01$. These smooth *EPR* changes embody the continuous nature of perceptual rivalry phase transitions.

Average $J_{ss}$ and *EPR* present a peak at $g=0.0103$ for $D$ values near 0.01, and around $D=0.1$, another peak appears at $g=0.0195$. For both situations, the system possesses three basins of nearly equal weight (Fig.8). This is attributed to the highest switching frequency and energy dissipation occurring when all three basins have equal weight. Additionally, ridges of *EPR* emerge in the *Down-Tri* and *Up-Tri* phase ranges. Evidently, phases with more states require more $J_{ss}$ flux for thermodynamic support, leading to a naturally higher cost for their complex behavior. Therefore, *EPR* offers a thermo-energetic perspective on global nonequilibrium phase transitions.

As discussed earlier, the *Down-Tri* phase is rarely observed in experiments (Fig.3(**c1**, **c2**)). Instead, increasing stimulation typically leads to direct switching from *Down-Single* to *Bi-Riv* phase, with *Up-Tri* emerging in some cases. *EPR* results support this: when $D$ is small, a single ridge appears in the *Down-Tri* phase, and its *EPR* values are significantly higher than other phases for larger $D$. The nonequilibrium thermo-energetic explanation is that the *Down-Tri* phase incurs the cost exceeding what the neural system can provide under normal circumstances. Distinguishing dim patterns in this phase requires more cost expenditure, incompatible with rapid perceptual alternations. This result predicts the need for additional energy input to sustain the *Down-Tri* phase transition. For the similar reason, reported experiments show that the *Up-Tri* phase exhibits unequal weights for *X*, *Y*, and *M* states, often concentrating on the mixed state *M* or unilateral states *X*, *Y* [12].

Interestingly, the ravine between two hills in Fig.8(**b1**, **b2**) consistently appears at around $g = 0.16$, independent of noise strength. This aligns with the average flux results, suggesting states remain confined to a single deep basin for extended periods, necessitating a high and steep barrier. Fig.6(**c1**, **c2**) confirms this expectation. Furthermore, binary rivalry (perceptual bistability) is a prevalent phenomenon in multiple visual and auditory rivalries. In this context, the principle of minimal cost, reflected by the minimal *HDR* and *EPR*, proposes a viable solution. This fact also suggests the relatively low irreversibility of the bistable state, as discussed in the next part. Besides, the *Down-Single* and *Up-Single* phases exhibit low *EPR* values similar to the *Bi-Riv* phase. This aligns with the expectation of low *EPR* for phases with only one basin.

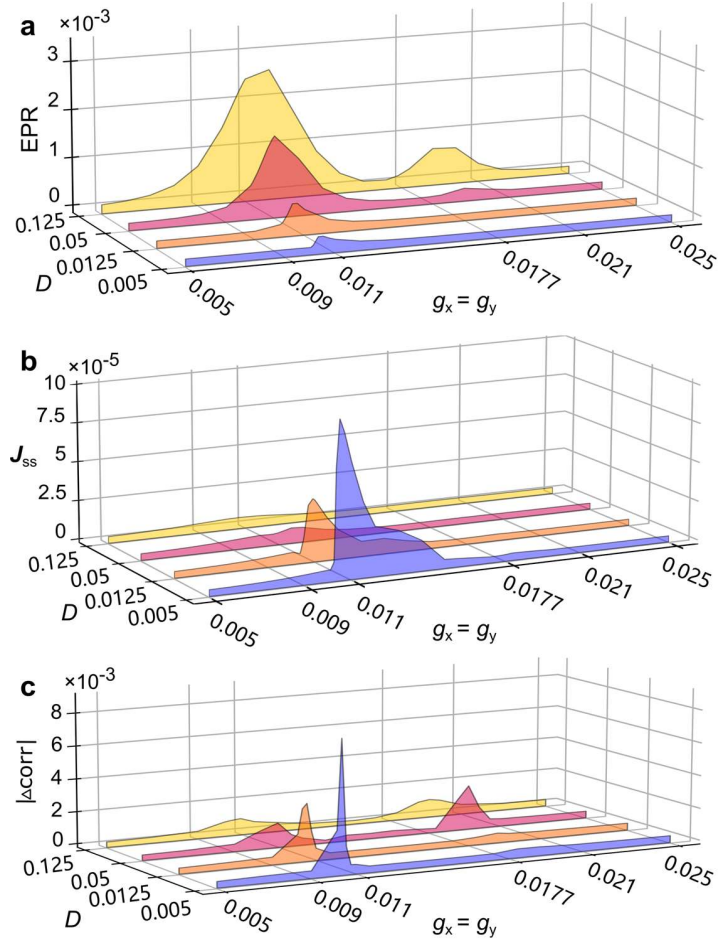

**Figure 9.** $|\triangle corr|$, *HDR* and the averaged $J_{ss}$

**Time irreversibility and detailed balance breaking.** As defined in the methods section, the time-irreversibility measure $|\triangle corr|$ quantifies the degree of detailed balance breaking. Fig.9 reveals a close correspondence between $|\triangle corr|$, average $J_{ss}$, and *EPR* trends. The directional nonequilibrium flux $J_{ss}$ drives detailed balance breaking, inducing system asymmetry, and average $J_{ss}$ captures the global Jss flux behavior, with higher values indicating more active dynamics and consequently higher cost reflected by *EPR*. This increased cost signifies a system further away from equilibrium. Thus, $|\triangle corr|$, average $J_{ss}$, and EPR naturally link together and exhibit similar patterns. Notably, average $J_{ss}$ and *EPR* are often challenging to measure directly, whereas $|\triangle corr|$ is readily obtained from finite time trace data using electrophysiological methods as discussed in methods section. This makes $|\triangle corr|$ a valuable tool for revealing the nonequilibrium thermodynamics and energetics.

$|\triangle corr|$ exhibits remarkable sensitivity. The optimal range for the *Down-Tri* phase ($D \in [0.005, 0.0125]$) reveals a prominent peak at $g = 0.0103$, precisely where all three basins have equal weight. Similar behavior is observed for the *Up-Tri* phase (Fig.9(**c**)). It is because that non-zero $|\triangle corr|$ indicates time-reversal asymmetry and detailed balance breaking. At low $D$ values, states are confined within one basin, resembling a monostable system. Conversely, at large $D$ values, the landscapes become flattened, and states fluctuate like white noise, neither exhibiting significant detailed balance breaking. This sensitivity suggests that $|\triangle corr|$ can pinpoint the practical parameter range relevant to biophysical function. This contributes to establish a connection between the results of experimental psychology and neurophysiology.

Importantly, finding a maximal or minimal value of $|\triangle corr|$ within the time sequence of perceptual switching can anticipate the occurrence of a phase transition. This is because such extrema correspond to significant slope changes in $|\triangle corr|$, $EPR$, and average $\boldsymbol{J}_{ss}$, which, as shown previously, reflect the onset of bifurcations or nonequilibrium phase transitions. This approach allows for direct prediction of these transitions from time traces, bypassing the need for complex dynamical modeling [88].

Furthermore, the *Down-Tri* phase exhibits much higher $|\triangle corr|$ compared to the *Up-Tri* phase (Fig.9(**c**)). In a view of the detailed balance breaking, it explains the experimental rarity of the *Down-Tri* phase that possesses a higher degree of nonequilibrium, while perceptual trisitability of *Up-Tri* phase is commonly reported recent years. In contrast, for the cases of monostability and bistability, the system is closer to the equilibrium. This reinforces our conclusion for the prevalence of *Bi-Riv* phase (perceptual bistability) in the experiments, beyond the analysis by average $\boldsymbol{J}_{ss}$ and $EPR$.

## Unequal stimulations and inhibitions

Real-world scenarios often involve unequal external stimuli (e.g., differing auditory intensities). We extend our framework to this general case ($g_x \neq g_y$) (Fig.10(**a1-a10**)). As expected, stronger stimulation of one side (e.g., $X$) leads to increased dominance. In *Down-Single* phase (initial $g_x = g_y$), increasing stimulation on $X$ leads to a new $X$-excited state emerging (Fig.10(**a1**)), signifying unilateral excitation. Further increase strengthens this dominance (obscuring previous bifurcations under $g_x = g_y$). This behavior is consistent across different $g_x \neq g_y$ values.

For cases with multiple exciting states (e.g., *Down-Tri*, *Bi-Riv*), increasing stimulation on $X$ weakens other excited states, including the mixed state (Fig.10(**a3-a8**)). When strong enough, $X$ becomes dominant with only faint traces of other excitations remaining (Fig.10(**a4**, **a6**, **a8**)). Interestingly, the mixed state (ground state in *Down-Tri* phase) persists in *Up-Tri* with relatively insufficient $X$ stimulation (Fig.10(**a3**, **a7**)).

The *Up-Single* phase exhibits unexpected behavior. Even with very strong $X$ stimulation, the mixed state remains (Fig.10(**a10**)). Further simulations (not shown) indicate this persists even at extremely high values ($g_x = 1$). This suggests another, even weak, stimulation exceeding a threshold can maintain its own excited state.

Fig.10(**a1-a4**) span *Down-Single*, *Down-Tri*, and *Bi-Riv* phases, demonstrating a consecutive process: the dominant perception emerges from a faint state and strengthens with increasing stimulation. This aligns with sensory decision-making processes observed in brain studies [76]. We infer similar behavior in brain decision-making for *Bi-Riv* and *Up-Tri* phases (Fig.10(**a4-a7**)). Under strong bilateral stimulation, a mixed state emerges, resembling neutral decision-making (*Up-Tri* phase). This suggests the emergence of a mixed state from a binary structure under strong stimulation may be a general phenomenon. Similar to Fig.10(**a7**, **a8**), strong unilateral stimulation can lead to the sequential decay of the weaker side and then the mixed state. These observations hint at potential homologies between various conflict and competitive behaviors in the neural system. These cases of unequal external stimulations support Levelt's Proposition I: stronger stimulation on one side increases its perceptual dominance. This is particularly evident in binary rivalry (Fig.10(**a4-a6**)).

The counterpart of the case $g_x \neq g_y$ is the situation of unequal inhibition $\beta_x \neq \beta_y$ (Fig.10(**b1**, **b3**, **b5**, **b7**, **b9**)), leading to a physiological imbalance between the inhibitory neural populations "X_inh." and "Y_inh." (Fig.1). For *Bi-Riv* and *Up-Tri* phases that are important for perceptions, Fig.10(**a3**, **a5**, **a7**, **a9**) and Fig.10(**b1**, **b3**, **b5**, **b7**) demonstrate that the effects of stronger stimulation on one side and stronger inhibition on another side produce similar effects. A proper fact that can be depicted by results of Fig.10 is the binocular imbalance in vision research. It is caused by two interrelated sources: Organic difference between two eyes corresponds to the case $g_x \neq g_y$ [89,90], and neural suppression at complex anatomical site [59,91-93]. Providing that the lateral geniculate bodies play the role as "X_inh." and "Y_inh." in Fig.1 [94,95], the inhibitory neural interactions of the binocular imbalance can be simulated and explained by regulating $\beta_x$ and $\beta_y$. Further, Fig.10(**b2**, **b4**, **b6**, **b8**, **b10**) manifest that this imbalance can be compensated largely by modulating the external stimulation $g_x$ and $g_y$ [58]. For a specific patient ($\beta_x = 0.0075$ and $\beta_y = 0.008$ in Fig.10(**b**)), $g_y - g_x = 0.00175$ indicates the "balance point" of external compensatory therapy [93]. In general, these methods and results can be employed in other types of imbalanced perceptions. For instance, in auditory system, considering that the module "Col_pop." represents the inferior colliculus and "X_inh.", "Y_inh." for nucleus of lateral lemniscus[96] (Fig.1), one can simulate the imbalance of auditory frequency sensitivity.

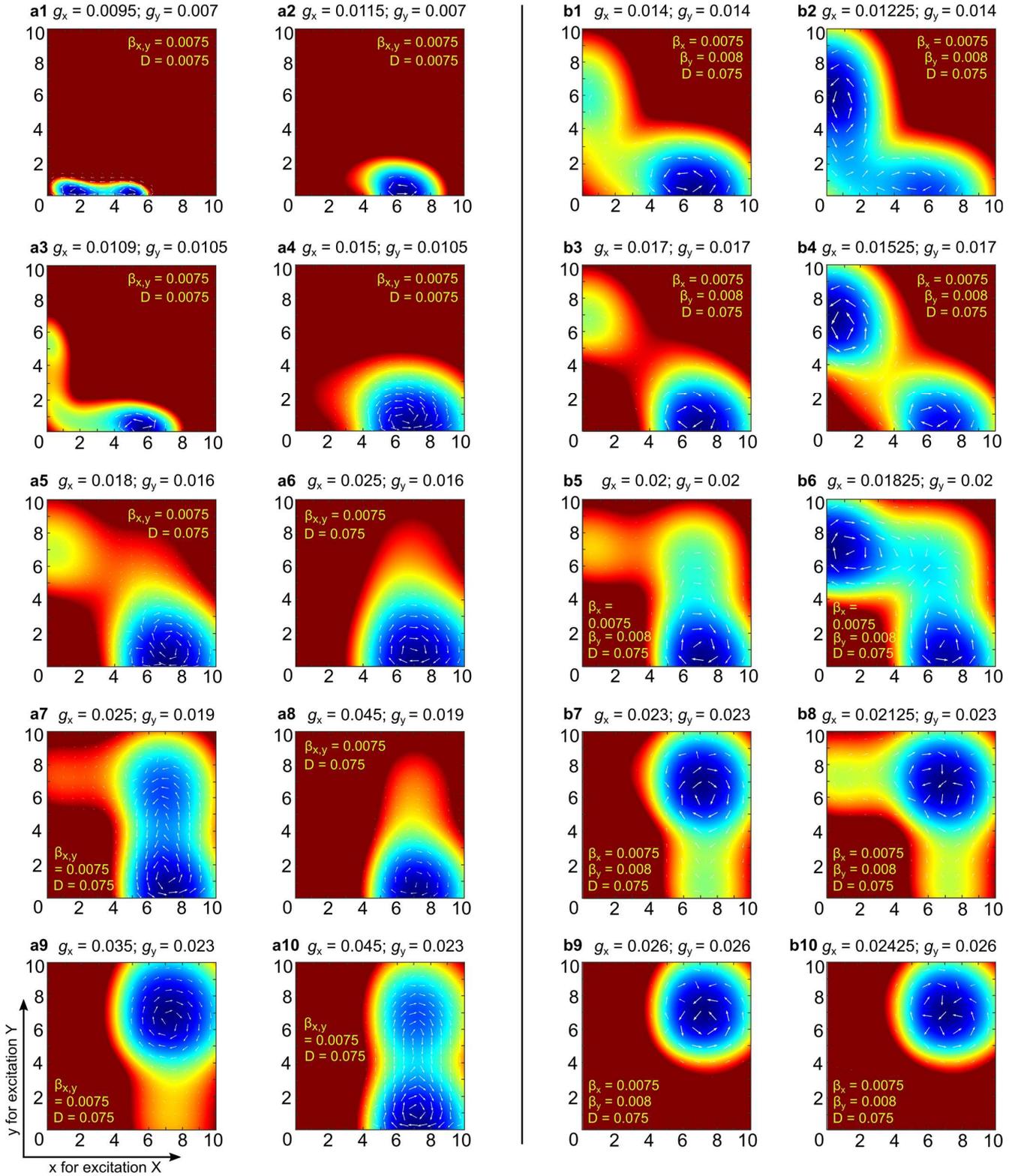

**Figure 10. Landscapes and corresponding vertical views in the condition** $g_x \neq g_y$ **and** $\beta_x \neq \beta_y$

Every graph of landscapes is in the same color range with Fig.2. Noticing that $x$, $y$ have the permutation symmetry, (a1-a10) actually contain all situations. (b1,3,5,7,9) exhibit the phase transition process from *Bi-Riv* to *Up-Single*, in which the inhibition of one side ($\beta_x$) is weaker than one of another side ($\beta_y$), while $g_x = g_y$. (b2,4,6,8,10) exhibit the cases corresponding to (b1,3,5,7,9), in which $g_x < g_y$ as a compensation, and the difference between $g_x$ and $g_y$ are constant.

# DISCUSSION

It can be saliently recognized that a neural exciting pool, "Col_pop.", exists and plays a critical role in our model, which accepts the external and internal feedback stimulations together, and then modulate and output the stimulation. In recent years, researchers have noticed the influence of neural pool on perception, especially that neural pool can promote the perception switching of various forms [97,98]. In fact, this neural pool plays a role as a concise hidden layer storing the information of the previous step in a recurrent neural network, while some reports have already studied the function of the memory in rivalry [99]. Meanwhile, as an instance, the binocular rivalry, which mainly happens in the primary visual cortex V1 and extrastriate cortex V2, V3 and V4, has also been modelled as the hierarchical neural network and some results have been achieved [100-102]. These findings indicate the connections between the hierarchical model, the recurrent neural network and the memory of nerve.

The mechanism, of which the perceptual neural population and neural pool are synchronously excited, then the inhibitory population coupled with the perceptual neural populations are stimulated by the neural pool, is crucial. It reflects some commonalities of various perceptual rivalries in neurophysiological anatomy. For visual neural pathway, lateral geniculate bodies play the role of the neural pool, delivering the ascending signals to V1 and receiving feedback from the higher cortex, meanwhile the inhibitory neurons located in the same receptive field are responsible for the projections of negative feedback [103,104]. Some studies report that a wide range of visual pathways are involved in the process of binocular rivalry [105,106]. For auditory pathway, the neural pool reflects the biofunction of inferior colliculus, while nucleus of lateral lemniscus and frequency-sensitive neural layers in inferior colliculus provide the inhibitory projection [96,107]. These largely entrench the neurophysiological rationality of our framework.

Recently, studies of the underlying dynamics reconstruction of the real biosystem provide a way to implementing our framework at a realistic level of neurophysiology [108]. Besides, some experimental paradigms, such as the tactile stimulation study in Ref.[3], involve multiple stimuli. These cases might necessitate expanding the parameter space within our framework, as detailed in Ref.[36]. Dimensionality reduction techniques like principal component analysis or adiabatic approximation could be employed for further model refinement [109]. Moreover, the development of biomimetic monomolecular catalysts and nano enzymes advances rapidly, which are found to be significant in the neural system these years [110-112]. The application of the nonequilibrium physics to enzyme kinetics was also carried out [113,114]. Thus, the nonequilibrium physics method that we introduce is also beneficial to studying neural science at the microscopic and mesoscopic scales.

Whereas, our framework is based on steady-state Markov process. For *Up-Tri* phase, this results in the equal probability of the paths $X \to M \to Y$ and $X \to M \to X$. As discussed in the section of dominant path, these results align with the study of brain decision making, but omit the effect of perceptual memory. This may be due to that the Markov processes integrate the adaptation term, erasing the influence of time delay and noise-induced perceptual resonance [71].

We conclude our work with a thorough comparison with the Levelt's propositions, for the reason as we have presented above that it has been extended to a wide range more than binocular rivalry, and many reports have provided tests and rectifications:

**Proposition I**: "Increasing the stimulus strength for one eye (one side of perceptions) will increase the perceptual predominance of that eye." While this proposition holds true as a fundamental tenet of perceptual rivalry, our unequal stimulation results reveal additional insights. Although increasing stimulus strength increases dominance, the weaker perception can remain its residual feature or even maintain a mixed state with extreme stimulation. This expands the original definition.

**Proposition II**: "Increasing the stimulus strength for one eye (one side of perceptions) will not affect the average perceptual dominance duration of that eye. Instead, it will reduce the average perceptual dominance duration of another eye." Our unequal stimulation results only partially support this proposition. Neither absolute nor relative average dominance durations remain constant, because increased stimulation deepens and expands the dominant basin, encroaching on the other basin and making it shallower and smaller. This supports recent debates challenging Proposition II [19,36] and aligns with research suggesting limitations [78]. Further work is needed to clarify whether this deviation reflects model limitations or inherent proposition constraints.

**Proposition III**: "Increasing the stimulus strength for one eye (perception) will increase the perceptual alternation rate." This proposition closely relates to Proposition II. For bistability, our unequal stimulation results diverge from the original proposition but align with recent studies [78]. The deeper basin of stronger stimulation holds the state for extended periods, with brief switches to the shallower basin.

**Proposition IV**: "Increasing the stimulus strength in both eyes while keeping it equal between eyes (perceptions) will increase the perceptual alternation rate." Similar to Proposition II, this proposition has been extensively debated. Recent research shows a decrease in alternation rate at low stimulation [78]. Our results for bistability also predict such deviation, contrary to the original proposition. However, the *Up-Tri* mixed state emerging with further stimulation offers a potential explanation. If observers interpret this stochastic bias towards one perception as individual perception, reported alternation rate may increase.

## NONEQUILIBRIUM PHYSICAL METHODS

For simulating the fluctuation in the neural system, we add white noise to the deterministic dynamics. For nerve cells, the sources of noise may be the fluctuation of ion concentration, the sensitivity of ion channel, or dissynchronization among neurons, etc. Therefore, the general Langevin equation, also the stochastic differential equations (SDEs) can be derived from Eq.(1):

$$\frac{dx}{dt} = F_1 + \sqrt{2D}\,\xi(t)$$

$$\frac{dy}{dt} = F_2 + \sqrt{2D}\,\xi(t) \quad (2)$$

In Eq.(2), $\xi(t)$ is the white noise with $\langle \xi(t)\xi(t')\rangle = \delta(t-t')$, and $D$ is the constant and isotropic diffusion coefficient, from diffusion matrix. From Eq.(2), the Fokker-Planck (FP) equation is derived as:

$$\frac{\partial P(x,y,t)}{\partial t} = -\frac{\partial}{\partial x}(F_1 P) - \frac{\partial}{\partial y}(F_2 P) + D\left(\frac{\partial^2}{\partial x^2} + \frac{\partial^2}{\partial y^2}\right)P \quad (3)$$

Eq.(3) can be rewritten as $\partial P(x,y,t)/\partial t = -\nabla \cdot \boldsymbol{J}$, in which $J_x = F_1 P - D(\partial P/\partial x)$ and $J_y = F_2 P - D(\partial P/\partial y)$. When the nonequilibrium system reaches a steady state, Eq.(3) becomes $\partial P/\partial t = -\nabla \cdot \boldsymbol{J} = 0$. In this case, the probability flux $\boldsymbol{J}_{ss}$ is rotational due to the divergent free condition $\nabla \cdot \boldsymbol{J}_{ss} = 0$, which does not vanish to zero and can be expressed as [115]:

$$\boldsymbol{J}_{ss} = \boldsymbol{F}P_{ss} - \boldsymbol{D} \cdot \nabla P_{ss} \quad (4)$$

$\boldsymbol{J}_{ss}$ demonstrates the flow of the probability in nonequilibrium steady state. Thus, the deterministic driving force $\boldsymbol{F}$ can be decomposed as $\boldsymbol{F} = \boldsymbol{J}_{ss}/P_{ss} - \boldsymbol{D} \cdot \nabla U$, where $U = -\ln(P_{ss})$ is the generalized potential landscape.

**Average $\boldsymbol{J}_{ss}$**. Whether for breaking the detailed balance or constituting the nonequilibrium thermodynamical driving force of the system, $\boldsymbol{J}_{ss}$ embodies its importance. We define the average $\boldsymbol{J}_{ss}$ to measure the strength of the flux as:

$$\int_{\Sigma} P_{ss}|\boldsymbol{J}_{ss}|^2 d\Sigma \quad (5)$$

where $\Sigma$ means that the integral traverses the whole phase plane.

**Entropy production rate and heat dissipation rate**. As a direct inference of the second law of thermodynamics, entropy production rate is significant to measure the energy exchange of the system. The definition of entropy is always employed as $S = -k_B \int_{\Sigma} P \ln P d\Sigma$ where $\Sigma$ is the whole domain of the system, and it contains two parts: The system's entropy $S_{sys}$ and the medium's entropy $S_{med}$. Thus, *EPR* equals to the sum of time change rates of $\dot{S}_{sys}$ and $\dot{S}_{med}$, reading:

$$T\dot{S}_{sys} = T\dot{S}_{tot} - T\dot{S}_{med} = \int_{\Sigma}(\boldsymbol{F} + \boldsymbol{D}\nabla U)\cdot \boldsymbol{J}\,d\Sigma - \int_{\Sigma}\boldsymbol{F}\cdot\boldsymbol{J}\,d\Sigma$$

$$= \int_{\Sigma}|\boldsymbol{J}|^2/P\,d\Sigma - \int_{\Sigma}\boldsymbol{F}\cdot\boldsymbol{J}\,d\Sigma$$

$$= EPR - HDR \quad (6)$$

Note that the Einstein's relation $D = k_B T$ is used in Eq.(6). Here, $T\dot{S}_{tot}$ reflects the difference of time change rates between the total work applied to the system and free energy reflected the averaged irreversible work. $T\dot{S}_{med}$ equals to the heat dissipation rate $\int_{\Sigma}\boldsymbol{F}\cdot\boldsymbol{J}\,d\Sigma$ (*HDR*) representing the time change rate of heat dissipated into the medium, which is also the ensemble average power exerted by $\boldsymbol{F}$ [116,117]. Therefore, determining $\boldsymbol{J}_{ss}$ enables immediate calculation of both EPR and HDR. Notably, the second law dictates $\dot{S}_{sys} = 0$ at steady state, implying that total entropy change equals to dissipated heat. Hence, analyzing *EPR* and *HDR* provides insights into system evolution and costs of the energy, work and dissipation, even without precise system energy measurement.

**Time irreversibility and nonequilibrium cross-correlation**. Since flux $\boldsymbol{J}_{ss}$ gives the origin of detailed balance breaking, we seek a measure to quantify its degree, which also indicates the level of time reversal symmetry violation. The absolute difference between the cross-correlation $\langle x(t)y(t+\tau)\rangle$ and its time-reversed counterpart $\langle y(t)x(t+\tau)\rangle$ serves this purpose, where $x$ and $y$ represent signals of two perceptions. Notably, this measure essentially integrates the transition probability between any two states ($A$, $B$) of the system under nonequilibrium steady-state conditions. As shown in previous studies [118,119], the equation for the difference $|\triangle corr|$ reads:

$$|\triangle corr| = \left| \int_\Sigma x^A P_{ss(t)}^A y^B P_{ss(t+\tau)}^B W^{AB(\tau)} d\Sigma - \int_\Sigma y^A P_{ss(t)}^A x^B P_{ss(t+\tau)}^B W^{BA(\tau)} d\Sigma \right|$$

$$= \left| \int_\Sigma xy [P_{ss}^A W^{AB(\tau)} - P_{ss}^B W^{BA(\tau)}] d\Sigma \right|$$

$$\cong \left| \int_\Sigma xy (\boldsymbol{J}_{ss}^{AB} \cdot \boldsymbol{e}_x + \boldsymbol{J}_{ss}^{AB} \cdot \boldsymbol{e}_y) \tau d\Sigma \right| \quad (7)$$

in which $W^{ij(\varepsilon)}$ is the propensity function representing the probability a state moves from state $i$ to $j$ during the time interval $\varepsilon$. Eq.(7) indicates that only when $\boldsymbol{J}=0$ does the time reversal symmetry and detailed balance hold with $|\triangle corr|=0$. Hence, larger $|\triangle corr|$ signify a greater departure from equilibrium and a stronger violation of time reversal symmetry. In essence, $|\triangle corr|$ quantifies the degree of irreversibility and statistically correlates positively with irreversible work. Therefore, in nonequilibrium steady state, $|\triangle corr|$ is valuable for predicting the trends of *EPR* and *HDR*, quantities often difficult to measure directly in experiments.

While Eq.(7) holds strictly for small time intervals $\tau$, practical limitations hinder exact integral calculations. Therefore, we use long time average instead of ensemble average. For parameter set $D$ and $g$, we calculate $|\triangle corr|$ for $x$, $y$ time series over a span of 400000. We select 201 time points as delay times, equally spaced between 0 and 200000. Each calculation is repeated 100 times and averaged. Finally, the results for the 201 delay times are summed and averaged again. This procedure yields the statistical characteristics of $|\triangle corr|$ and enables its application to finite sequences, directly applicable to the finite-time trace data obtained through neurophysiological methods like EEG, ECoG, fMRI, and invasive electrophysiological measurements [52-55,57,120].

**Dominant Path**. Based on the steady probability distribution, we can identify the most probable path connecting two arbitrary points, which defines the dominant path. In classical mechanics, this is achieved by processing the Lagrangian by the Euler-Lagrange equation. For stochastic dynamics, the Lagrangian is replaced by the Onsager-Machlup action [121]:

$$\mathcal{L}_{OM} = \frac{1}{4D}[\dot{\boldsymbol{x}} - \boldsymbol{F}(x,y)]^2 + \frac{1}{2}\boldsymbol{\nabla}\cdot\boldsymbol{F}(x,y) \quad (8)$$

The probability of a certain path is:

$$P[X(t=0\sim\tau)] = \exp\left(-\int_0^\tau \mathcal{L}_{OM} dt\right) \quad (9)$$

However, analytical solution of the dominant path refers to the two points boundary value problem (TPBVP), which is often challenging. Besides, the dominant path for connecting two points does not always possess a global maximal probability. Hence, we directly discretize the Onsager Machlup function and derive its first derivative [122-125], then employ genetic gradient annealing algorithm to minimize the Onsager Machlup action of a path connecting two stable nodes of different states.

# Appendix

In Eq.(2) of the deterministic force, coefficients are set up as:

$\gamma=0.1$, $\sigma=0.01$, $\alpha=0.00275$,

For every case except Fig.10(**b1-b10**), $\beta_x=\beta_y=0.0075$;

In the part $\Lambda(x)$ and $\Lambda(y)$ for the adaptation, $a_1=0.15$, $a_2=1$, $c=5$;

In $\phi_{col\_inh}$, $q=8.0557\times 10^{-3}$; In $\phi_{self\_inh}$, $\eta=5.04\times 10^{-4}$;

And for the sigmodal input-output function, $\theta=5\times 10^{-3}$, $k=2.25\times 10^{-3}$.

**ACKNOWLEDGMENTS**
Y.W. and L.X. was supported by the National Science Foundation of China (Grants No.21721003) and the Natural Science Foundation of Jilin Province, China (Grant No.20210101141JC).

[87] The essential cause derives from the definiting equations Eq.(5, 6). In the testing calculation, we find that the integral of the identical part $|\boldsymbol{J}_{ss}|^2$ of Eq.(5, 6), can almost be regarded as a constant, and the flux $\boldsymbol{J}_{ss}$ will only observely distribute in the basins of a landscape. Moreover, the global $P_{ss}$ for every calculations in this work are normalized to 1. Thus, when $D$ is small, the narrow and deep basins possess the mostly probability of the whole landscape, and each cell of the integral Eq.(5) has a relatively large $|\boldsymbol{J}_{ss}|^2$ and $P_{ss}$ values. When $D$ is large, the opposite is true. In other words, Eq.(5) calculates the average value of $|\boldsymbol{J}_{ss}|^2$ in the basins based on a normolized global $P_{ss}$. The result is larger if the $|\boldsymbol{J}_{ss}|^2$ value in each cell of the integral is larger, and vice versa. The similar analysis can be used to interpret the trend of EPR along with $D$ increasing.